\documentclass{article}

\usepackage{arxiv}

\usepackage[utf8]{inputenc} 
\usepackage{hyperref}       
\usepackage{url}            
\usepackage{booktabs}       
\usepackage{amsfonts}       
\usepackage{nicefrac}       
\usepackage{microtype}      
\usepackage{amsmath}
\usepackage{cleveref}       
\usepackage{lipsum}         
\usepackage{graphicx}
\usepackage{natbib}
\usepackage{doi}

\title{A 72h exploration of the co-evolution of food insecurity and international migration}

\date{June 28, 2024 (Text revision: July 2, 2024)}

\newif\ifuniqueAffiliation
\uniqueAffiliationtrue

\ifuniqueAffiliation 
\author{
    \href{https://orcid.org/0009-0001-1376-024X}{\includegraphics[scale=0.06]{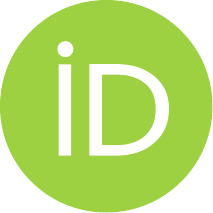}\hspace{1mm} Duncan Cassells} \\
    LIP6, Sorbonne Université\\
    médialab, Sciences Po\\
    LPI, Université Paris Cité\\
    Paris, FR\\
    \texttt{duncan.cassells@lip6.fr}\\
    \And
    \href{https://orcid.org/0000-0002-8346-1907}{\includegraphics[scale=0.06]{orcid.pdf}\hspace{1mm} Lorenzo Costantini} \\
    CENTAI Institute\\
    Turin, IT\\
    \texttt{lorenzo.costantini@centai.eu}\\
    \And
    \href{https://orcid.org/0009-0004-4356-5949}
    {\includegraphics[scale=0.06]{orcid.pdf}\hspace{1mm}Ariel ~Flint Ashery} \\
    Department of Mathematics\\
    City, University of London\\
    London, UK\\
    \texttt{Ariel.Flint-Ashery@city.ac.uk}\\
    \And
    \href{https://orcid.org/0009-0004-2700-2003}
    {\includegraphics[scale=0.06]{orcid.pdf}\hspace{1mm}Shreyas Gadge} \\
    Institute for Biodiversity\\ and Ecosystem Dynamics, \\University of Amsterdam, NL\\
    \texttt{s.r.gadge@uva.nl}\\
    \And
    \href{https://orcid.org/0000-0002-6069-7474}
    {\includegraphics[scale=0.06]{orcid.pdf}\hspace{1mm}Diogo L. ~Pires} \\
    Department of Mathematics\\
    City, University of London\\
    London, UK\\
    \texttt{Diogo.L.Pires@city.ac.uk}\\
    \And
    \href{https://orcid.org/0009-0007-3870-546X}{\includegraphics[scale=0.06]{orcid.pdf}\hspace{1mm}Miguel Á. ~Sánchez-Cortés} \\
    Faculty of Information Engineering, Computer Science and Statistics\\
    Sapienza University of Rome\\
    Rome, IT\\
    \texttt{sanchezcortes.2049495@studenti.uniroma1.it}\\
    \And
    \href{https://orcid.org/0009-0007-3870-546X}{\includegraphics[scale=0.06]{orcid.pdf}\hspace{1mm}Arnaldo ~Santoro} \\
    Department of Environmental Sciences, Informatics and Statistics\\
    Ca' Foscari University\\
    Venezia, IT\\
    \texttt{arnaldo.santoro@unive.it}\\
    \And
    \href{https://orcid.org/0000-0002-6748-5124}{\includegraphics[scale=0.06]{orcid.pdf}\hspace{1mm}Elisa ~Omodei} \\
    Department of Network and Data Science\\
    Central European University\\
    Vienna, AT\\
    \texttt{omodeie@ceu.edu}\\
}
\else
\usepackage{authblk}

\newbox{\orcid}\sbox{\orcid}{\includegraphics[scale=0.06]{orcid.pdf}} 
\author[1]{%
	\href{https://orcid.org/0000-0000-0000-0000}{\usebox{\orcid}\hspace{1mm}David S.~Hippocampus\thanks{\texttt{hippo@cs.cranberry-lemon.edu}}}%
}

\author[1,2]{%
	\href{https://orcid.org/0000-0000-0000-0000}{\usebox{\orcid}\hspace{1mm}Elias D.~Striatum\thanks{\texttt{stariate@ee.mount-sheikh.edu}}}%
}
\affil[1]{Department of Computer Science, Cranberry-Lemon University, Pittsburgh, PA 15213}
\affil[2]{Department of Electrical Engineering, Mount-Sheikh University, Santa Narimana, Levand}
\fi

\hypersetup{
pdftitle={A 72h exploration of the co-evolution of food insecurity and international migration},
pdfsubject={q-bio.NC, q-bio.QM},
pdfauthor={David S.~Hippocampus, Elias D.~Striatum},
pdfkeywords={First keyword, Second keyword, More},
}

\begin{document}
\maketitle

\begin{abstract}

Food insecurity, defined as the lack of physical or economic access to safe, nutritious and sufficient food, remains one of the main challenges of the 2030 Agenda for Sustainable Development. Food insecurity is a complex phenomenon, resulting from the interplay of environmental, socio-demographic, and political events. Previous work has investigated the nexus between climate change, conflict, migration and food security at the household level, however these relations are still largely unexplored at national scales. 
In this context, during the Complexity72h workshop, held at the Universidad Carlos III de Madrid in June 2024, we explored the co-evolution of international migration flows and food insecurity at the national scale, accounting for remittances, as well as for changes in the economic, conflict, and climate situation. To this aim, we gathered data from several publicly available sources (Food and Agriculture Organization, World Bank, and UN Department of Economic and Social Affairs) and analyzed the association between food insecurity and migration, migration and remittances, and remittances and food insecurity. 
We then propose a framework linking together these associations to model the co-evolution of food insecurity and international migrations.
\end{abstract}



\section{Introduction}


Achieving food security (i.e., economic and physical access to sufficient, safe, and nutritious food for everyone everywhere - \cite{WorldFoodSummit1996}) is an ambitious target 
embedded in Sustainable Development Goal 2~(\cite{nations2015transforming}). Unfortunately, in 2023, nearly 282 million people across 59 countries or territories still experienced severe food insecurity and needed food assistance~(\cite{WFP})
Seminal papers investigate factors affecting food security at the household level. Among these factors, low income and education, climate change, and conflicts feature as characteristics of food insecure households~(\cite{allee2021cross,d2023drivers,smith2017assessingFood,smith2017world}).

These drivers are also common to international migration phenomena~(\cite{castelli2018drivers}). Food insecurity has been reported as a mediating factor of such drivers on migration~(\cite{morales2020exploringConnections}) and both its presence and inequality in regards to it are correlated to out migration~(\cite{smith2022food}). In turn, migration impacts food insecurity both at destination and origin country in several ways. For example, being an immigrant increases one's likelihood to be food insecure~(\cite{smith2017assessingFood}). On the other hand, even though migration can lead to labour loss at the origin country, it also decreases the quantity of food required at household level~(\cite{zezza2011assessingImpact}). 
Furthermore, international remittances sent by migrants account for a relevant fraction of several country's international investments, especially in low- and middle-income countries where their flows are larger than official development assistance~(\cite{ratha2018migrationWorldBank}). In those cases, migration may lead to an increase of household income. This in turn improves food security through better food consumption and other indirect investments~(\cite{zezza2011assessingImpact,obi2020international}). This has been analysed at national level, where total remittances are associated with total migration fluxes and the difference between origin and destination country income per capita~(\cite{ratha2010outlook}).

In general, seminal works explore the relation between migration and food security focusing on household level data retrieved through surveys~(\cite{allee2021cross,d2023drivers,smith2017assessingFood,smith2017world}). 
Against this background, we present a set of preliminary analyses exploring the linkage between food security and international migration, considering indicators at the national scale. 
In this way, we aim to contribute to advancing the understanding of the nexus between these phenomena, eventually allowing for the development of effective polices for ensuring food security across countries.


\section{Materials and Methods}

In this article we employ three main publicly available datasets coming from different sources: the Food and Agriculture Organization of United Nations (FAO), the World Bank Group (WBG), and the United Nation Population Division (UNPD). We use FAO data to obtain national food insecurity indicators (\cite{DATASETfaofoodindicators}). In particular, we obtain the $3$-year average percentage of country population affected by \emph{severe food insecurity}.
Moreover, from this dataset, we also obtain other annual national indicators: Temperature Change on Land, Gross Domestic Product per capita (GDP), and Political Stability, quantified by the Political Stability and Absence of Violence/Terrorism Index (\citep{DATASETfaotemperaturechange,DATASETfaopopulation}). Temperature Change on Land measures annual mean temperature change with respect to a baseline temperature measured during the 1951–1980 period, while the Political Stability index measures ``perceptions of the likelihood of political instability and/or politically-motivated violence, including terrorism" (\cite{stabilitydefinition}). This index gives each country's score in units of a standard normal distribution, ranging from $-2.5$ to $2.5$. Furthermore, we obtain yearly received remittances from the WBG data, containing compensation and money transfers between countries converted in current USD (\cite{DATASETwbreceivedremittances}), along with each country's income group label (low income, lower-middle income, upper-middle income, and high income) as assigned by the WBG in 2019. Finally, from the UNPD, we obtain open data on international migration (\citep{DATASETmigrationprocessed, DATASETmigrationraw}), i.e. estimates of the international migrant stock for the mid-point (1 July) for each available year.

\begin{figure}[!ht]
	\centering
	\includegraphics[width=0.9\textwidth ]{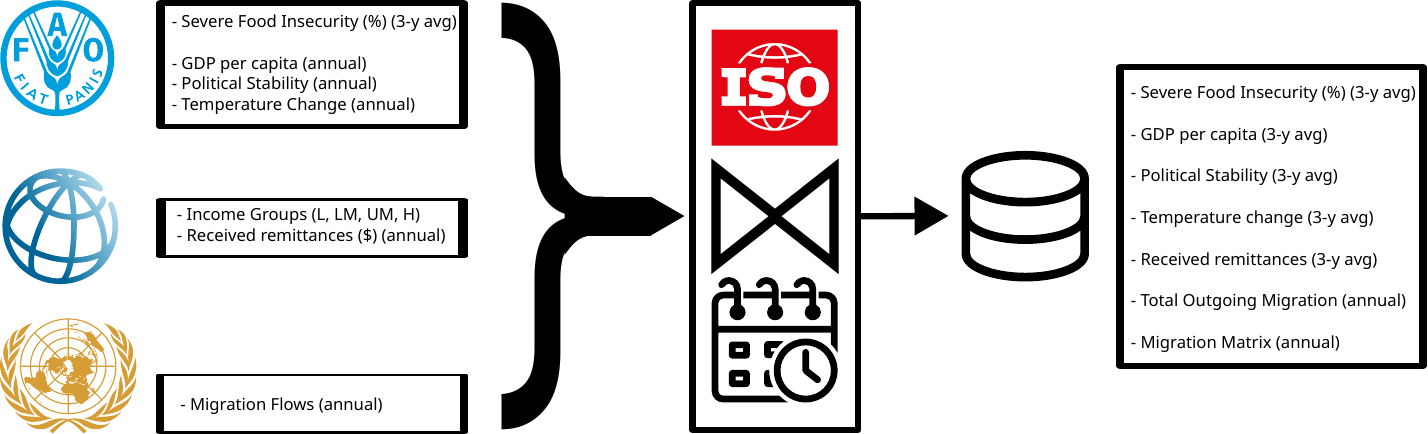}
	\caption{\textbf{Data cleaning and aggregation flow}. Data from FAO, WBG, and UNPD are combined by means of their ISO3 country codes and made consistent by computing 3-year averages from annual indicators.
    To produce this figure, we employed Creative Commons icons from \cite{thenounproject} by the authors Rikas Dzihab and New icon.}
	\label{fig:dataflow}
\end{figure}

To ensure consistency throughout the study and to be able to compare each yearly indicator (temperature change, GDP, political stability, etc.) with food insecurity, we computed the $3$-year averages for all yearly indicators. 
At the same time, when combining the datasets, we considered the ISO3 codes for each considered country as used by FAO and by the WBG (\citep{DATASETfaoiso3, DATASETwbiso3}), and filtered the dataset to consider only countries that had available food insecurity and GDP data. The data pipeline followed to obtain the dataset for this study is illustrated in Figure~\ref{fig:dataflow}.

\section{Results}
During the Complexity 72h workshop, we carried out a preliminary analysis of the complex nexus between food insecurity and international migration. The initial analyses here presented are organized in three macro-steps linking: (i) food insecurity and out migration, (ii) migration flows and remittances, and (iii) the combination of the main drivers of food insecurity (i.e., the economic, conflict and climate situation) with remittances to investigate food insecurity. In the following subsections we present the initial results obtained during the workshop. 

\subsection{Food insecurity and international migration}
\begin{figure}[!ht]
	\centering
 \includegraphics[width=0.75\textwidth ]{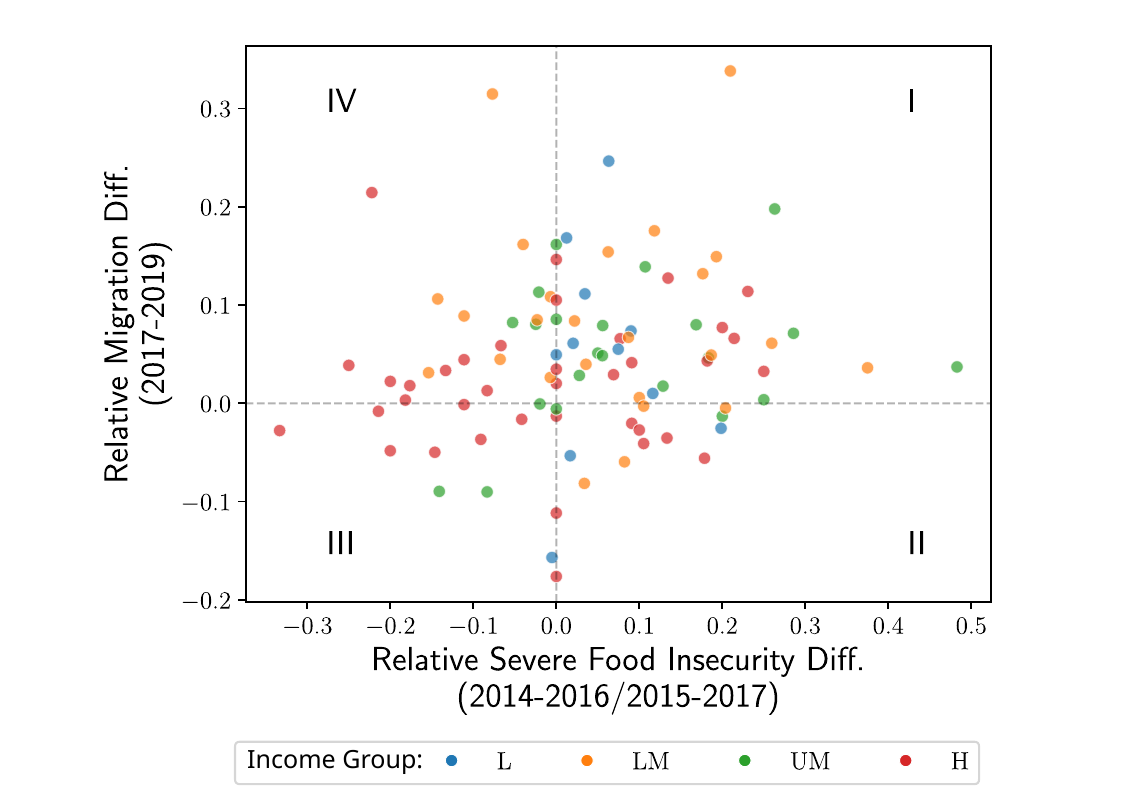}
 \caption{\textbf{Relative changes in future total outgoing migration in function of relative changes in the origin country food insecurity situation}. Each point represents a country and is coloured by its income group: Low (L), Lower-Middle (LM), Upper-Middle (UM), High (H). 
 }
	\label{fig:insecurity_migration}
\end{figure}

Food insecurity is one of the drivers of international out migration.
Thus, we hypothesize that relative changes in total international out migration from a country in the future might positively correlate with relative changes in the current food insecurity situation in that country, as suggested by~\cite{smith2022food}. We empirically explored this association in our data as shown in Figure~\ref{fig:insecurity_migration}. 
Although the current limited data did not allow us to infer a statistically significant correlation between the two variables,
qualitatively, we notice that numerous countries are in the first and third quadrants. Countries in the first quadrant have recorded a relative increment in food insecurity and a relative increment in future migration rate. Conversely, in the third quadrant, where food insecurity decreases, migration out-fluxes in the next years decrease as well. 

\subsection{International migration and remittances}
\label{sec:migration and remittances}
Better earning opportunities are one of the drivers of international migration. These allows migrants to contribute to the livelihood strategies of their households through \emph{remittances}, i.e. the money they send to the household~(\cite{zezza2011assessingImpact}. With the upsurge of international migration, remittances have expanded dramatically and are linked to economic development and poverty reduction~(\cite{zezza2011assessingImpact}. They are also seen as a significant capital flow to developing countries~(\cite{ratha2010outlook}). 

In this study, we use a simplified version of the model for estimating bilateral remittances suggested in~\cite{article}. The model assumes remittance flow between countries to be affected by three factors: the migrant stocks in different
destination countries, incomes of migrants in the different destination countries, and, to some extent,
incomes in the origin country. The average remittance sent back home by a migrant from country $i$ living in destination country $j$ is modelled as a function of the per capita income of the origin country and the destination country:
\begin{equation}
r_{ij} = f(Y_{i},Y_{j}) = 
\begin{cases}
Y_i, \quad& \text{if } Y_j < Y_i;\\
Y_j, \quad& \text{if } Y_j \geq Y_i.\\
\end{cases}
\label{eqn:rij}
\end{equation}
Here, $Y_{i}$ is the average per capita GDP of the origin country and $Y_{j}$ is the average per capita GDP of the destination country. The rationale is that the migration occurs in the expectation of earning a higher level of income for the dependent household than what the migrant would earn in their home country. So, it is assumed that the average remittance by an individual is at least as much as the per capita income of the home country, even when the individual migrates to a lower-income country. 

The total amount of remittances received by country $i$ is then proposed to be estimated as:
\begin{equation}
\hat{R_{i}} = \sum_{j} r_{ij} M_{ij},
\label{eqn:remittance}
\end{equation}
where $M_{ij}$ is the migration vector between country $i$ and different destination countries $j$. 

\cite{article} proposed the relation between the level of remittance ($r_{ij}$) and the per capita income of origin and destination countries ($Y_{i}$, $Y_{j}$) to have an exponential parameter $\beta$. But we assume a simplification with $\beta = 1$, which gives us Equation~\ref{eqn:rij}. Moreover, in contrast to the use of average per capita GNI for $Y_{i}$ and $Y_{j}$, we use the average per capita GDP, to keep consistent variables across our different analyses. 

We use Equations ~\ref{eqn:rij} and ~\ref{eqn:remittance} to estimate the remittance values for all origin countries  using migration and average GDP data. Then, we compare the estimated remittances $\hat{R_{i}}$ with actual remittance data $R_{i}$ for the period 2017-2019. 
\begin{figure}[!ht]
	\centering
 \includegraphics[width=0.65\textwidth ]{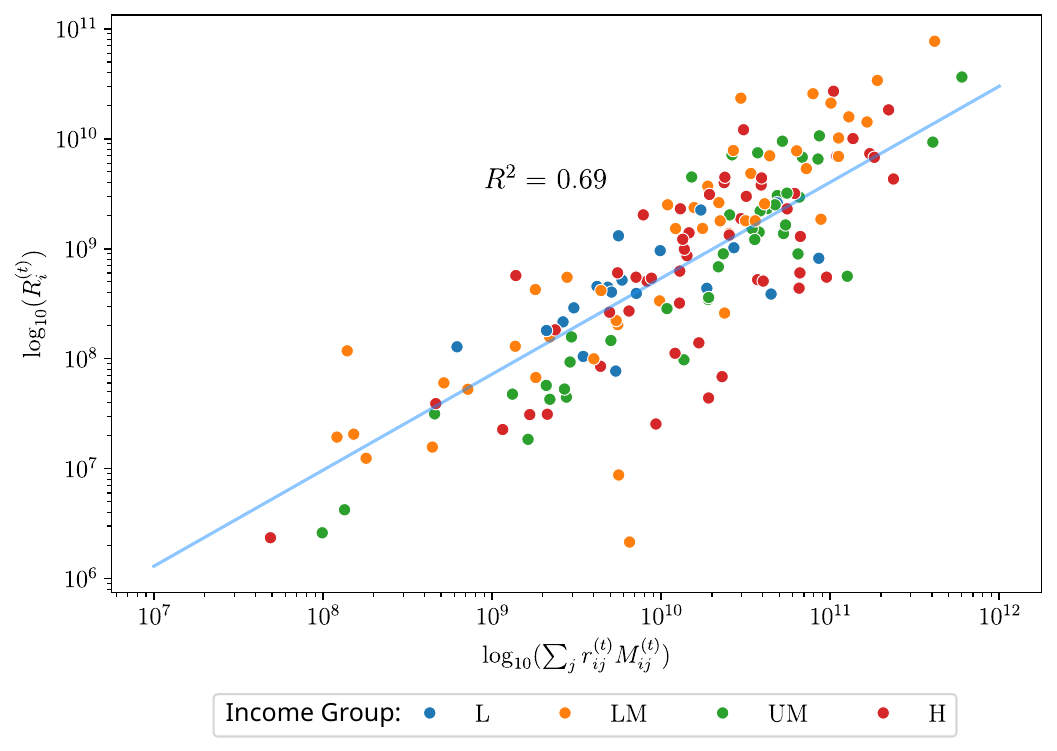}
 
 \caption{\textbf{Relationship between the observed received average remittances of each country for the 2017-2019 period and the estimated average remittances for the same time period}. Each point represents a country and is coloured by the its income group: Low (L), Lower-Middle (LM), Upper-Middle (UM), High (H).}
	\label{fig:remittances}
\end{figure}
\newline
In Figure \ref{fig:remittances} we observe a linear relation on the log-log scale for the estimated vs observed remittances plot with a slope value of $\gamma \simeq 0.87$. This implies a modification to Equation ~\ref{eqn:remittance} to take the form:
\begin{equation}
\hat{R_{i}}^{(t)} =  \left(\sum_{j}r_{ij}^{(t)}M_{ij}^{(t)}\right)^\gamma
\label{eqn:modified remittance}
\end{equation}
This validates the assumptions of the model by \cite{article} and gives a modified relation between remittance received by origin country and the migrant stock from that country. This relation can be used in modeling remittances from international migration flows to explore their effects on food insecurity. 

\subsection{Remittances and other drivers of food insecurity}
Food insecurity is driven by a variety of factors affecting food availability, access, and utilization.
Its main drivers are related to the political stability of the country, the conditions of its national economy, and the effects of climate change. Naturally, agricultural production is tied to environmental change. Conflicts, related to political instability, diminish investments in agricultural technology and the capability of the state to provide infrastructure for food storage, distribution and access. Drought and other climate shocks are aggravated by conflict, which further complicates the impacts on food security and livelihood. The status of a national economy is subject to a multitude of other forces, including climate and internal politics, inequality, and population pressures. The status of the national economy has a feedback effect on food insecurity. In Section~\ref{sec:migration and remittances}, we discussed the effects of remittance on poverty reduction and economic development, which could be linked to food insecurity. 

In this study, we decompose the evolution of food insecurity into two components: international migration, mediated by a function $G(R)$ of remittances $R(M)$, and internal factors related to economic well-being in the home country (GDP per-capita, $e$), environmental change (estimated using soil temperature variation, $T$), and political stability ($s$). 
Furthermore, we hypothesize (for simplicity, as an initial exploration) that we can write the relative change in food insecurity at the national level as a linear combination of the following variables: 
\begin{equation}
    \frac{f_i^{t+\Delta t} - f_i^{t}}{f_i^{t}} = \alpha + \beta \log_{10} \left(\frac{R_i^t}{R_i^{t-\Delta t}}\right) + \gamma\frac{s_i^{t} - s_i^{t-\Delta t}}{s_i^{t-\Delta t}}+ \zeta(T_i^{t} - T_i^{t-\Delta t})+\eta\frac{e_i^{t} - e_i^{t-\Delta t}}{e_i^{t -\Delta t}}
\label{eqn:delta food insecurity}
\end{equation}
where the logarithmic functional form for the remittance ratio was chosen on the basis of an initial empirical exploration of the association between relative changes in food security and received remittances.

We estimate the coefficients of Equation \ref{eqn:delta food insecurity} through multiple linear regression, using the method of ordinary least squares. The value of each independent variable at time $t$ is calculated from the data using a three-year rolling-average. Regression is performed using data from the yearly period 2014-2016, with $\Delta t$ corresponding to a one-year shift in the period start and end year. Results are reported in Table~\ref{tab:regression} and used as an initial placeholder for the broader modeling framework proposed in the next section. However, the signal is statistically weak and further analyses on additional data are required to draw solid conclusions.

\begin{table}
\centering
\begin{tabular}{lclc}
\toprule
\textbf{Dep. Variable:}    & $(f_i^{t+\Delta t} - f_i^{t})/f_i^{t}$  & \textbf{Model:}                &  Ordinary Least Squares   \\
\textbf{  R-squared:}      &                        0.236                        & \textbf{  Adj. R-squared:    } &     0.201   \\
\textbf{ F-statistic:}     &                        6.720                        & \textbf{  Prob (F-statistic):} &  9.24e-05 \\
\textbf{BIC:}              &                        -86.81                       & \textbf{  AIC:               } &    -99.42   \\
\textbf{Log-Likelihood:}   &                        54.709                       &                                &             \\
\bottomrule
\end{tabular}

\vspace{0.5cm}

\begin{tabular}{lcccccc}
\toprule
                & \textbf{coefficient} & \textbf{std err} & \textbf{t} & \textbf{P$> |$t$|$} & \textbf{[0.025} & \textbf{0.975]}  \\
\midrule
$\alpha$  &       0.4436  &        0.143     &     3.112  &         0.003        &        0.160    &        0.727     \\
$(e_i^{t} - e_i^{t-\Delta t})/e_i^{t-\Delta t}$      &      -1.2307  &        0.658     &    -1.871  &         0.065        &       -2.538    &        0.076     \\
$(s_i^{t} - s_i^{t-\Delta t})/s_i^{t-\Delta t}$      &       0.0450  &        0.020     &     2.265  &         0.026        &        0.006    &        0.085     \\
log($R_i^t/R_i^{t-\Delta t}$) &      -0.2103  &        0.132     &    -1.598  &         0.114        &       -0.472    &        0.051     \\
$T_i^{t} - T_i^{t-\Delta t}$      &      -0.1230  &        0.033     &    -3.746  &         0.000        &       -0.188    &       -0.058     \\
\bottomrule
\end{tabular}
\vspace{0.25cm}
\caption{
\textbf{OLS regression results} linking relative changes in remittances and in other known food insecurity drives (i.e., economic situation, conflict, and climate) with future relative changes in food insecurity levels.}
\label{tab:regression}
\end{table}

\begin{figure}[!ht]
	\centering
	\includegraphics[width=0.65\textwidth ]{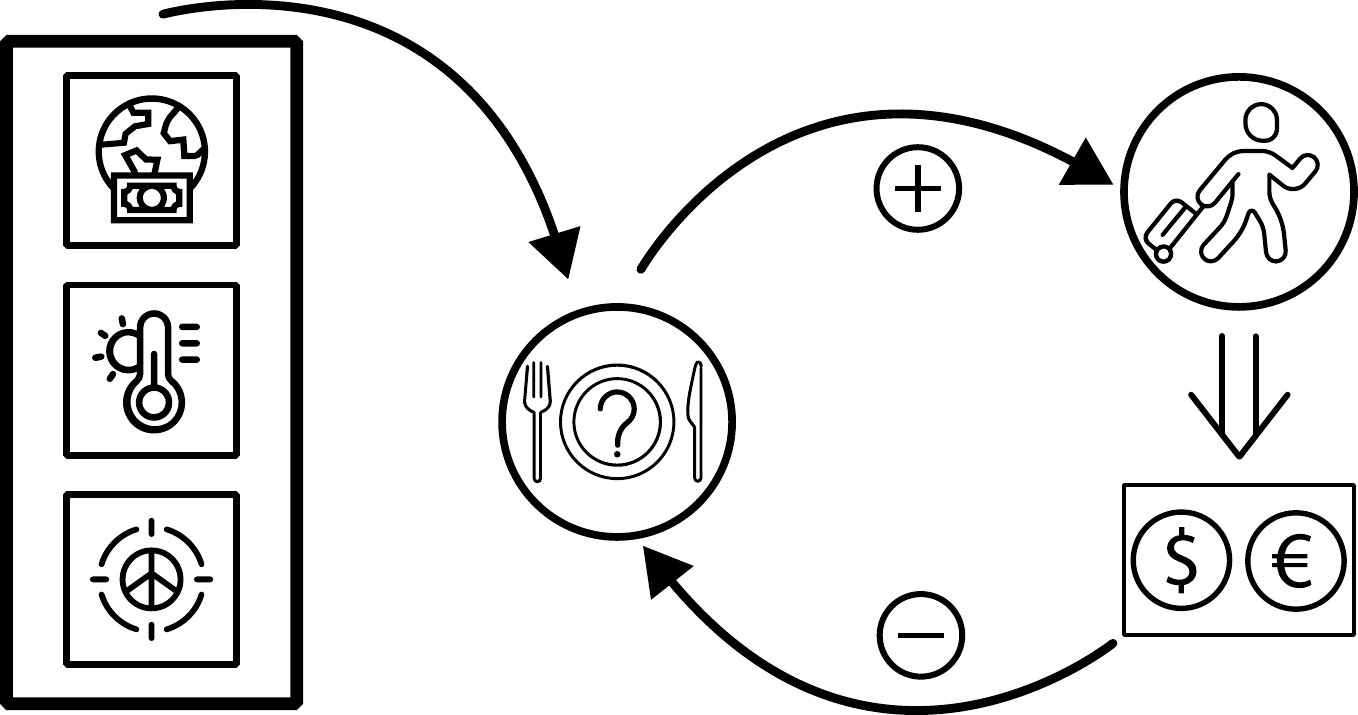}
    \caption{\textbf{Schematic representation of the proposed framework to investigate the co-evolution of food insecurity and international migration}. 
    Food insecurity can drive  individuals to leave their home country. Once people reach more secure countries, they tend to send remittances to relatives who remained in their home country, which can help them reduce their food insecurity. Finally, we propose to estimate changes in food insecurity in the origin country  considering the known drivers of food insecurity (i.e., economic situation, conflict, and climate) as well as the positive effect of remittances (mediating the impact of outgoing migration).
    To produce this Figure, we employed Creative Commons icons from \cite{thenounproject} by the authors: Sergey Novosyolov, Moch Rizki Eko Waluyo, UNKNOWN, bsd studio, Geni Alando, and yuni sarah.}
	\label{fig:model_sketch}
\end{figure}

\subsection{Exploring an integrated model for the co-evolution of food insecurity and international migration}
In light of the preliminary analyses just presented, we argue that there is room to develop a mechanistic model to address the food security-migration nexus. In this direction, we propose a preliminary model for the co-evolution of food insecurity and international migration which combines the three investigations described above to obtain an analytical framework linking food insecurity, migration, remittances, and the other food insecurity drivers, in accordance to the diagram presented in Figure \ref{fig:model_sketch}. 
Developing a conceptual framework of this kind could be useful to study the possible co-evolution of future migration fluxes, remittances, and food insecurity at the national scale.

In this context, we preliminarily consider the impact of food insecurity on international migration (in terms of country's total outgoing migration) as a linear function of the following form:
\begin{equation}
    \frac{m_i^{t+\Delta t} - m_i^{t}}{m_i^{t}} = \alpha + \beta \frac{f_i^{t} - f_i^{t-\Delta t}}{f_i^{t-\Delta t}}
    \label{eqn:Food2Migration}
\end{equation}
where $m_i^t$ is total number of people from country $i$ living abroad at time $t$, $f_i^{t}$ is food insecurity of country $i$ at time $t$, and $\alpha$ and $\beta$ are parameters to be estimated from the data (fitting the available data, we get $\alpha = 0.0436$ and $\beta = 0.0846$).

\begin{figure}[!ht]
\centering\includegraphics[width=.75\textwidth ]{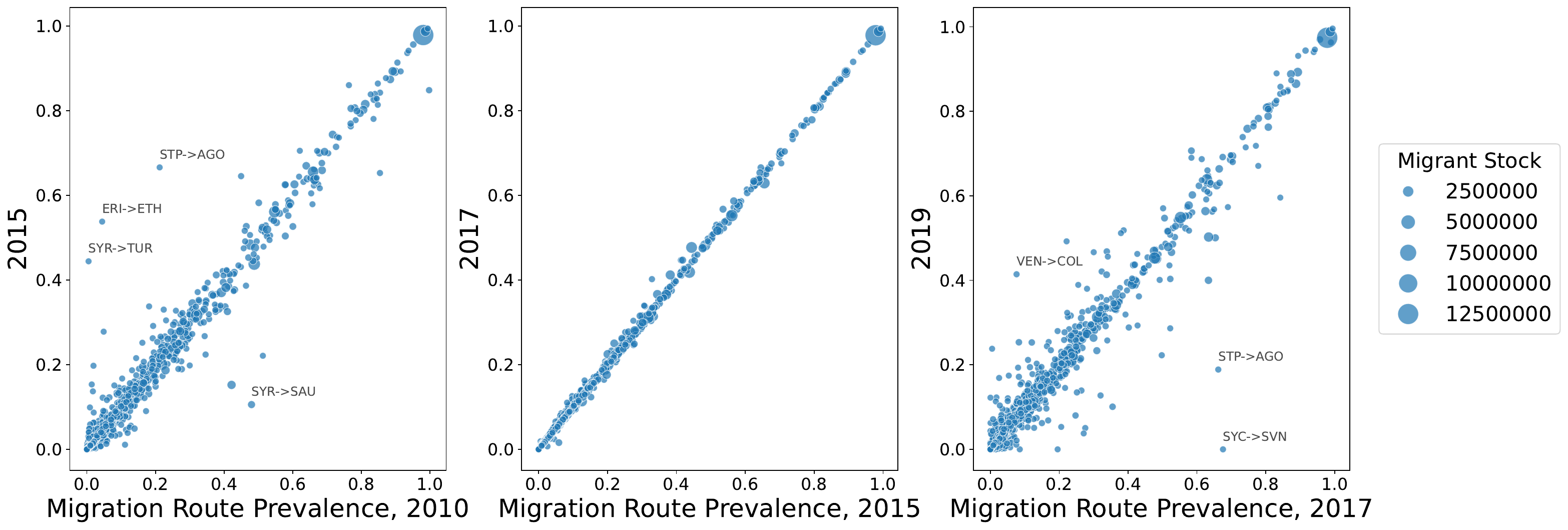}
 \caption{\textbf{Migration route prevalence}. Migration route prevalence is the migrant stock that is in one country as a proportion of the total migrant stock existing worldwide from the origin country. By comparing migration route prevalence at a time $t$ with a time $t+\Delta t$, we may observe variations in destination choice over time. All three comparisons display a strong correlation between two subsequent time periods, motivating our assumption that the likelihood of destination choice can be assumed to be static over time.}
\label{fig:migration_deltas}
\end{figure}

Once the total number of people deciding to migrate out of a country is determined, we make the assumption that the profile of the destination countries is kept unchanged throughout the period of time further considered. Figure~\ref{fig:migration_deltas} shows an empirical validation of this hypothesis in our data. 

As presented in Section \ref{sec:migration and remittances}, the initial study linking international migration and remittances showed that the relation between the two could be described by Equation \ref{eqn:modified remittance}. Based on this, we estimate the total remittances $\hat{R}_{i}^t$ received by a country $i$ at each time $t$ through Equation \ref{eqn:modified remittance} and using the parameter $\gamma=0.8731$ obtained from the fitting of the data.

Finally, food security can be updated considering remittances, political stability, soil temperature variation, and national per-capita GDP following Equation \ref{eqn:delta food insecurity}. The five fitted coefficients reported in Table~\ref{tab:regression} were used in the model.

The preliminary model developed considers the relationships discussed above and described in Figure \ref{fig:model_sketch}. This was initialized with country level data on food insecurity and total migration from 2017 to 2019.
We used the data available on GDP per capita, temperature change and political instability both for the initializing period of 2016 to 2019 and the simulation period. Due to the time constraints of the workshop under which this work was performed, we leave the analysis of simulations estimating changes in food security and  migration flows up to 2023 for the future. Following preliminary simulations and their analysis, we will extend this to different altered scenarios where the impact of temperature changes and economic shocks will be considered.

\section{Discussion}
During the Complexity 72h workshop, we performed an exploratory analysis of the links between food insecurity and international migration.
In this preliminary study, we qualitatively observed a general positive trend between the changes in food insecurity and out-going migration flows, thus motivating one side of our model's logic.
The other part of the model assumes that migrants living abroad send remittances back home. On this side, we proposed a preliminary approach to estimate the remittances that migrants send to the origin country. Our results show that 69\% of the variance in the remittance data is accounted for by migration flows and differences in, providing initial evidence for the model.
Finally, we evaluated through a simple linear model how remittances (mediating migration flows), changes in temperature at soil, per-capita GDP, and political instability are associated with changes in food insecurity.

The initial analysis presented in this report presents several limitations and ample room for improvements. These lie both with the data and the model itself.

Using data from three different sources presents challenges in matching time period coverage between them. Migration stock data is available every 5 years from 1990 to 2015, as well as for 2017 and 2019; while food insecurity data is available as aggregates of three consecutive years starting for the period 2014-16 and updated every year; and socioeconomic data for income groups, received remittances, GDP per capita, and political stability is available every year for the whole time period.
Further complication arises from inconsistent collection for countries between the three sources and naming convention inconsistencies for countries themselves - index keys to match between sources were implemented to achieve consistent results, but the impact of missing countries was not explored due to time constraints of the workshop.
Further data exploration and gathering will provide more complete data records that would increase confidence in our evaluations.

Modelling limitations arise from restricting the scope to the co-evolution of migration and food insecurity (i.e., we did not attempt to model how these affect the economic situation of countries, climate change, and political instability), and lack of extensive data to evaluate the results.
In an attempt to uncover the links among the considered variables, we explored linear relationships between food insecurity and migration flows, and how remittances, climate chance, economic situation, and political instability may affect food insecurity. Higher order relationships were not considered.
No evaluation of the model is presented due to the limited time of the workshop and data availability. 
Hence, this report only intends to provide a preliminary data exploration and conceptual formalization to model this significant problem.

Future development on this exploration will aim to build robustness and confidence in the framework and output. Alternative data sources may be a source of longer time span and improved internal consistency of each country's indicators. Future modelling work will develop more complex analytical frameworks to deepen investigation into the relationships amongst the underlying variables, also considering possible effects of several control variables. 

\section{Acknowledgements}
This report is the output of the Complexity72h workshop, held at the Universidad Carlos III de Madrid in Leganés, Spain, 24-28 June 2024. https://www.complexity72h.com. M. Á. acknowledges the Sicomoro Foundation for financial support to carry out this project.

\bibliographystyle{unsrtnat}
\bibliography{references}  






\end{document}